&latex209

\documentclass[preprint2]{aastex}

\shorttitle{The compact HII region}
\shortauthors{Roman-Lopes et al.}

\begin{document}


\title{The stellar population associated with the IRAS
 source 16132-5039 
\footnote{Based on observations made at Laborat\'orio Nacional de Astrofisica/MCT, Brazil}}


\author{A. Roman-Lopes, Z. Abraham}
\affil{Instituto de Astronomia, Geof\'\i sica e Ci\^encias Atmosf\'ericas, Universidade de
S\~ao Paulo \\ Rua do Mat\~ao 1226, 05508-900, S\~ao Paulo, SP, Brazil}

\email{roman@astro.iag.usp.br}


\begin{abstract}
We report the discovery of a young massive stellar cluster and infrared nebula
in the direction of the CS molecular cloud associated to the 
IRAS point source 16132-5039. 
The analysis of the mid-infrared images from the more accurate MSX catalog,
reveled that there are two independent components associated with the IRAS source. 
The integral of the spectral energy distribution for these components,  
between 8.28 $\mu$m and 100 $\mu$m, gave  lower limits for the bolometric luminosity
of the embedded objects of 
$8.7 \times 10^4 L_\odot$
and $9 \times 10^3 L_\odot$, which 
corresponds to  ZAMS O8 and B0.5 stars, respectively. 
The number of
Lyman continuum photons expected from the stars that lie along the reddening line 
for early-type stars
is about 1.7 $ \times$ 
10$^{49}$ s$^{-1}$, enough to produce the detected flux densities at 5 GHz.
The NIR spectrum of the nebula 
increases with frequency, implying that free-free
emission cannot be the main source of the extended luminosity, from which we conclude that the observed
emission must be mainly dust scattered light.
A comparison of the cluster described in this paper with the young stellar cluster associated with the IRAS source
16177-5018, which is located at the same distance and 
direction, shows that the mean visual absorption of the newly discovered cluster is about 10 magnitudes smaller and
it contains  less massive stars, suggesting that it was formed from a less massive
molecular cloud.

\end{abstract}




\section{Introduction}

Massive stars are born within dense molecular clouds forming clusters and associations; 
during their formation and early evolution they are often  visible only at infrared and radio wavelengths. 
With the advent of the large bidimensional near-infrared array detectors, the morphological 
and photometric studies of extremely young galactic stellar clusters have benefited. At 
near-infrared wavelengths (1 to 2.5 $\mu$m) it is possible to probe deep into the dense dust clouds where star 
formation is taking place.  

The strong ultraviolet (UV) radiation emitted by the massive stars dissociates and ionizes the gas, forming
compact HII regions seen at radio 
wavelengths. A large fraction of the  radiation could also
heat the dust, which eventually radiates  in 
the far-infrared (FIR); for this reason,
compact HII regions are among the brightest and most luminous objects in the 
Galaxy at 100$\mu$m. 
 
Because massive stars evolve very fast, they are extremely rare and difficult to find, except when related 
to the emission of the surrounding molecular cloud. In that sense, CS and NH$_3$ lines at 
radio frequencies, characteristic  of high 
density gas,  are good  tracers of massive star forming regions.
 
In this paper we present observations in the near infrared,  of a young stellar cluster of massive stars 
in the direction of  the IRAS point source I16132-5039. This work is a part of a survey aimed 
to the identification of stellar populations in the direction of IRAS sources that have colors characteristics
of ultracompact HII regions \citep{wood89} and strong CS (2-1) line emission \citep{bronf96}. 
The studied region is located in the  direction of another massive star forming region associated to the
IRAS point source 16177-5018 \citep{rom03}; they are part of the RCW 106 complex, 
located in the southern Galactic plane at a distance of 3.7 kpc \citep{cas87}.
The near-infrared cluster presented here had recently been located  by Dutra et al. (2003) by visual inspection 
of the 2MASS images; however their work contains only the possible cluster location, and
as Persi, Tapia \& Roth (2000) already showed  for NGC6334IV,  a high concentration of K band sources
can be due to localized low extinction and be mistaken with a stellar cluster, leading to
false identifications.

The cluster is associated with the radio source 332.541-0.111,  which presents continuum 
radio emission at 5 GHz as well as hydrogen recombination lines \citep{cas87}. 
The observations and data reduction
are described in section 2, the results are presented in section 3 and our main conclusions are
sumarized in section 4.

\section{Observations and data reduction}

The imaging observations were performed in June 2001 and May 2003 with the Near Infrared Camera (CamIV) of 
Laborat\'orio Nacional de Astrofisica (LNA), Brazil,  equipped with a Hawaii 1024x1024 pixel 
HgCdTe array detector mounted on the 0.6 m Boller \& Chivens telescope.The observations consisted of  8'x 8' frames in 
the direction of the IRAS source 16132-5039; the plate scale was 0.47 
arcsec/pixel and the mean values of the PSF full width at half maximum (FWHM) were 1.4, 1.7 and 2.1 arcsec
at the $\it{J}$, $\it{H}$ and $\it{nbK}$ images.
The total integration time was 2220 s for $\it{J}$, 1440 s for $\it{H}$ 
and 6400 s for $\it{ nbK}$ filters, resulting in a sensitivity at 3$\sigma$ of 18.2, 17.4 and 14.2 magnitudes
and completeness limits of 17.8, 16.5 and 13.8 magnitudes respectively. 
Details about the  calibration and reduction procedures can be found in Roman-Lopes et al. (2003).

Photometry from 2MASS All Sky Point Source Catalogue \footnote{http://www.ipac.caltech.edu/}
in the $\it{J}, \it{H}$ and $\it{K_S}$ filters became available recently, with  completeness limits of 
15.8, 15.1 and 14.3 magnitudes, in the three passbands \citep{egan01}.
Since the 2MASS $K_{S}$ band photometry, centered at 2.17 $\mu$m and with a bandpass of 0.32 $\mu$m has 
a completeness limit greater than our nb$K$ photometry, we completed our
observations with
the 2MASS catalogue astrometry and magnitudes in this filter. 
The comparison of our photometry with the 2MASS survey data, in an area of about 40 square
arcmin, corresponding to 
a total of 469, 610 and 440  common sources in the $\it{J, H}$ and 
$\it{K}$ bands, respectively,  is shown in figure 1.  We see a good linear relation between the two systems, with a slope of 1 and
a dispersion that increases with the magnitude.

\section{Results \& Discussion}

The combined false-colour infrared image, ($\it{J}$ blue, $\it{H}$ green and nb$\it{K}$ red)
 of the whole field is displayed in Figure 2, together with amplified individual images of the nebular region 
 at all bands. 
All of them, but especially  the $\it{H}$ and nb$\it{K}$ images, shows the presence of a small  spheroidal
nebula with a bright star at its center.

\subsection{The IRAS source}

In Figure 3 we present a contour map constructed from the LNA nb$K$ image, with a beam size of $2\times 2$ pixels
which shows the region around IRAS16132-5039. The contours start at 
2.2$\times 10^{-4}$ Jy/beam and are spaced by this same value.
The IRAS coordinate  has an intrinsic error  delimited by the ellipse 
plotted in the figure.   
A more accurate position for the IR source was obtained from the
Midcourse Space Experiment - MSX point source catalog $(psc)$ \footnote{
http://www.ipac.caltech.edu/ipac/msx/msx.html}. The MSX surveyed the entire 
Galactic plane within 
$\mid b \mid \leq 5^\circ$ in four mid-infrared spectral bands centered at 8.28, 
12.13, 14.65 and 21.34 $\mu$m, with image resolution of 19 arcsec and a global absolute astrometric accuracy of about
1.9 arcsec \citep{price01}. 
We found one MSX source, with coordinates $\alpha(\rm{J2000)=16^{h}17^{m}02.47^{s}}$, 
$\delta(\rm{J2000)=-50^{d}47^{m}03.5^{s}}$
within the IRAS uncertainty ellipse; its coordinates coincides with the star we
labeled IRS1.

However, looking at the MSX images, we found another closeby source, although outside the IRAS uncertainty ellipse,
with coordinates $\alpha(\rm{J2000)=16^{h}16^{m}55.94^{s}}$, $\delta(\rm{J2000)=
-50^{d}47^{m}07.8^{s}}$.
In Figure 4 we present our $H$ band image  overlaid with the 8.28$\mu$m band MSX contour diagram, with 
the contours spaced 
by $ 8\times 10^{-6}$ W m$^{-2}$ sr$^{-1}$, starting at $2.8\times 10^{-5}$ W m$^{-2}$ sr$^{-1}$. 
The same source was found
in the contour plots from the other MSX bands.
We designated the stronger
source as A and the other as B. While the IRS1 object and the infrared nebulae
are related to source A , 
source B  is also
associated with a small nebular region that shows a concentration of NIR
sources, as can also be seen in Figure 4. 

Since only the 8.28 $\mu$m flux density was given in the MSX $psc$ for both sources,
we integrated the flux density of each individual source for the four mid-infrared bands;  the
results are presented in table 1, which also shows the values from  MSX and
IRAS catalogs. Our integrated flux density for source B in the 8.28 $\mu$m coincides within 5\% with
the value given in the MSX catalogue, but it is 30\% larger for the A source. This discrepancy can
be understood if we consider that the automatic MSX algorithm subtracted part of the source B flux density
as background contribution. The relative contribution of source B decreases with increasing wavelength,
explaining also why it was not resolved by the MSX algorithm. It should be noticed that the reported IRAS flux
density at 12 $\mu$m coincides with the sum of our derived flux densities at 12.13 $\mu$m within 10\%,
while the MSX value was about 50\% lower.

In Figure 5 we plotted the mid to far-infrared spectral energy 
distribution of the sources A and B, without any correction for absorption.
We assumed that the IRAS flux density in the far infrared is divided between the two
sources in the same way as in the mid-infrared (14.65 and 21.33 $\mu$m): about 90\%  originating
from  source A and 10\%  from source B.
We then integrated the observed flux densities in the mid-far infrared, assuming 
a distance of 3.7 kpc, obtaining a luminosity $L_{\rm A}=8.7 \times 10^4 L_\odot$
and $L_{\rm B}=9 \times 10^3 L_{\odot}$ for sources A and B, respectively. Assuming that
the IR flux density represents a lower limit for bolometric luminosity of the embedded stars, 
we derived ZAMS  spectral types O8 and B0.5   for the  sources A and B, respectively  (Hanson et al. 1997).

\subsection{Cluster population}

In order to examine the nature of the stellar population in the direction of
the IRAS source, we analyzed the stars in two delimited regions:
one that we labeled "cluster", which contains the nebula and another that we 
labeled "control", which has a stellar population that we believe is dominated
by "field" stars, as illustrated in Figure 6. 
We represented the position of all objects detected in the $H$ 
band by crosses and we can see that the small region labeled "cluster" shows a concentration of sources.
In figure 7 we present comparative ($J-H$) versus ($H-K$) diagrams for the stars
detected in  the  $J$, $H$ and $K$ images,
together with the position of the main sequence, giant branch and reddening
vectors for early and late type stars (Koornneef 1983). 

We see that the stellar population of the "cluster" region is quite different
from that of the "control" region, with many sources lying on the right side of
the reddening vector for early type stars, showing excess emission at 2.2 $\mu$m.
In the "control" region the majority of the sources have colors of reddened
photospheres, with many objects located along the reddening vector for late
type stars.
It is well established that very young pre-main sequence objects present large infrared excess due 
to the presence of warm circumstellar dust (Lada \& Adams 1992).
Our results suggest that the stellar population in the direction of the IRAS source 
is very young, as can be inferred from their position in the $(J-H)$ versus $(H-K)$ diagram.

We separated the cluster sources from the field stars, 
by selecting all sources that lie to the right  or on the reddening vector 
for early type stars in the cluster's color-color diagram. 
Table 2 shows the coordinates and photometry  of all selected sources ($J$, $H$ and 
$K$ magnitudes from both LNA and 2MASS surveys).

Further information about the nature of the selected objects in Table 2 can 
be extracted 
from $J$ versus $(J-H)$ color-magnitude diagram shown in Figure 8.
We used this diagram instead the $K$ versus $(H-K)$ color-magnitude
diagram  to minimize the efect of the "excess" of emission
in the NIR on the derived stellar spectral types.
The locus of the main-sequence 
for class V stars at 3.7 kpc (Caswell \& Haynes 1987) is also  plotted, with the position of each spectral 
type earlier than A0 V indicated. 
The intrinsic colors were taken from Koornneef (1983) while 
the absolute $J$ 
magnitudes were calculated from the absolute visual luminosity for ZAMS taken 
from Hanson et al. 
(1997). 
The reddening vector for a ZAMS B0 V star, taken from Rieke $\&$ Lebofsky (1985), is shown by 
the dashed line with the positions of  visual extinctions  $A_V = 10$ and 20 magnitudes  
indicated by filled circles. We also indicated the sources with and without "excess"
in the color-color diagram by open and filled triangles respectivelly.

For the assumed distance, we estimated 
 the spectral type of the
stars that do not present excess emission in the near infrared, 
by following the de-reddening vector in the color-magnitude diagram, for the others we only 
gave a rough classification; the results are shown in the
last column of Table 2.  The main source of error in the derived spectral types arises from
uncertainties in the assumed cluster distance, which was derived from the velocities of the
radio hydrogen recombination lines. Since the closest distance was used, the main uncertainties
come from the errors in the galactic rotation curve model and related parameters. 
From the works of Blum et al. (1999, 2000, 2001) and Figuer\^edo et al. (2002) comparing kinematic
with spectroscopic distances we find that they
do not differ in more than 1 kpc, in which case the change 
in the luminosity class would be of two sub-spectral types for early O stars and one for early B stars.

We must notice that source IRS1, which is
associated with  MSX source A, has an estimated spectral type of at least
O5, reddened by about $A_{V} \approx$ 14 magnitudes. Besides, 
there are five objects (IRS3, IRS11, IRS18, IRS21 and IRS33) associated
with the mid-infrared source B; IRS3 
is probably an  O8 ZAMS star reddened
by about $A_{V}$= 7 magnitudes, while the others have estimated spectral types of early-B stars. 
The corresponding  bolometric luminosities are in agreement with the lower limits 
derived from the integrated  
mid-far infrared flux densities, corresponding to O8 and B0.5, for sources A and B respectively, 
as seen in section 3.1.


A lower limit to the number of Lyman-continuum photons 
produced in the star forming region, can be calculated taking into account only the
stars that do not show "excess" in the 
color-magnitude diagram (IRS1, 6, 8, 9, 10, 13, 14, 16, 20, 23, 26, 27 
and 35 in Table 2). It was computed from
the relation  given by Hanson et al. (1997), resulting in
$1.7 \times 10^{49}$ photons s$^{-1}$. Is interesting to note that
IRS1 is responsible for more than 90\% of the Lyman continuum
photons, being the main ionization source in the whole region.

It is also possible to obtain  the number of ionizing Lyman continuum photons $N_{Ly}$ from the radio
continuum flux density given by Caswell \& Haynes (1987), using the expression derived by Rubin (1968):
\begin{equation}
N_{Ly}= \frac{5.59\times 10^{48}S(\nu) D^2 T_e^{-0.45} \nu^{0.1}}{1+f_i[<He^+/(H^++He^+)]} 
\end{equation}
where  $\nu$ is in units of 5 GHz and $f_i$ is the fraction of  helium recombination photons that 
are energetic enough to ionize hydrogen. 
Using the values of $T_{e} = 4500$ K, $S(\nu) = 3.3$ Jy and $D$= 3.7 kpc
taken from Caswell \& Haynes (1987), we find that 
$f_i\approx 0$
and $N_{Ly} \sim 6 \times 10^{48}$ photons s$^{-1}$, compatible 
with the lower limit derived from the observed stars.

\subsection{The Infrared Nebula}

We can see from the detailed  $J,$ $H$ and nb$K$ images in figure 2 that the nebula
presents a spheroidal shape, approximately symetric around the 
IRS1 source.
In figure 9 we shown contour maps of the  nebular region at $\it{J, H}$ and nb$\it{K}$ bands obtained from our
infrared images. The contours were calibrated in flux with the values starting at 
0.44 ($J$), 1.6 ($H$) and 
2.2$\times 10^{-4}$ Jy/beam ($K$), with intervals of 0.37, 1.8 and 1.1$\times
10^{-4}$ Jy/beam respectively.
We estimated the total flux density by measuring the area
between contours and multiplying by the value of the corresponding  
flux density per unit area, obtaining $S(J)=0.023$ Jy, $S(H)=0.07$ Jy and
$S(K)=0.11$ Jy. We then corrected these results for extinction, using the
mean value $<A_{V}>$ = 15.1 magnitudes, derived from the stars in the direction of the
nebula that do not present infrared excess (IRS1, IRS6 and IRS10), and the standard extinction
law  from Rieke \& Lebofski (1985). We obtained
$S(J)=1.16$ Jy, $S(H)=0.8$ Jy and $S(K)=0.52$ Jy.
 
In the previous section we showed that the number of ionizing photons available from the
detected stars is enough to explain the radio continuum flux density measured by 
Caswell \& Haynes (1987). 
We will investigate now the contribution of free-free emission
to the observed nebular IR flux density.
Assuming constant density and temperature across the cloud and  local thermodynamic
equilibrium, the flux density $S(\nu)$ due to free-free emission can be written as: 
\begin{equation}
S(\nu)=\tau_\nu B_\nu(T)\Omega
\end{equation}
where $\tau_\nu$ is the optical depth at frequency $\nu$,  $B_\nu(T)$ is the Planck function
\begin{equation}
B_\nu(T)=\frac{2h\nu^3}{c^2}\frac{1}{{\rm exp}(h\nu/kT)-1}
\end{equation}
and 
$\Omega$ is the solid angle of the source given by:
\begin{equation}
\Omega=\pi{(L/2D)}^2
\end{equation}
where $L$ is the diameter of the ionized cloud and $D$ the distance to the observer.
The optical depth at a given frequency $\nu$ is:
\begin{equation}
\tau_\nu=\alpha_{\rm ff}L
\end{equation}
where $\alpha_{\rm ff}$ is the free-free absorption coefficient $\rm (cgs)$ taken from Rybick (1979):
\begin{equation}
\alpha_{\rm ff}=\frac{3.7\times 10^8\;[1-{\rm exp}(-h\nu/kT)]\;n_e n_i\;g_{\rm ff}(\nu,T)}{\nu^3\;Z^{-2}\; T^{1/2 }}
\end{equation}
where $g_{\rm ff}(\nu,T)$ is the Gaunt factor  obtained from Karzas \& Latter (1961).

For two frequencies $\nu_1$ and $\nu_2$ the ratio of the corresponding flux densities $S(\nu_1)$
and $S(\nu_2)$ may be calculated from:
\begin{equation}
\frac{S(\nu_1)}{S(\nu_2)}={exp}[h(\nu_2-\nu_1)/kT]\frac{g_{\rm ff}(\nu_1,T)}{g_{\rm ff}(\nu_2,T)}
\end{equation}

For a flux density in the 5 GHz continuum of 3.3 Jy and an electron 
temperature $T_e = 4500$ K, 
as given by Caswell \& Haynes (1987) the expected flux densities at 
the $J, H$ and $K$ bands are 0.05, 0.10 and 0.16 Jy, respectively. 
We verify that the measured values are much larger than what
was expected from the free-free emission, derived from the radio data. In fact, only an absorption
as low as 4 magnitudes would explain the inferred spectrum as free-free emission. 
Since in the direction of the A source, this absorption
is incompatible with even the less absorbed star (IRS1), we believe that this is not the 
main source of extended IR emission.

Accepting the mean value of $<A_{V}>$ = 15.1 magnitudes as the mean 
value for the absorption in the direction of A source, 
we found that the corrected flux density  increases with  frequency,
 suggesting that the observed extended radiation is
 scattered light from the nearby stars. 
We adjusted then a black body to the NIR  
fluxes, obtaining a good fit for T$\approx$ 16000K,  characteristic of middle-B  stars (Hanson
et al. 1997). 
 Lumsden $\&$ Puxley (1996), analyzing the ultracompact HII region G45.12+0.13, also
obtained an extinction corrected flux density that increases with decreasing wavelengths and
interpreted it as due to stellar light, scattered by dust through the HII region.

\subsection{Conclusions}

Near-IR imaging in the direction of the CS molecular cloud associated with the 
IRAS source 16132-5039, revealed an embedded young massive
stellar cluster. We detected 35  member 
candidates up to our completeness limit, concentrated in an area of about 2 square parsec. 
All images, but especially  the $\it{H}$ and nb$\it{K}$ bands, show the presence of a small  spheroidal
nebula with a bright star (IRS1) at its center.

The stars associated with the IRAS point source were identified using more accurate 
positions from the MSX catalogue. The analysis of the mid-infrared images
reveled that there are two  sources associated with IRAS 16132-5039. 
The strongest coincides with  the position of at least a dozen of OB stars, while
the weaker source is associated to less massive objects, with 
spectral types characteristic of
middle-B ZAMS stars.
The integral of the spectral energy distribution of the  
MSX-IRAS sources, between 8.28 $\mu$m and 100 $\mu$m, gives lower limits to the bolometric luminosity
of the embeded objects of 
$L=8.7 \times 10^4 L_\odot$
and $L=9 \times 10^3 L_\odot$, which 
corresponds to  ZAMS O8 and B0.5 stars, respectively (Hanson et al. 1997). 
The results are compatible with the spectral types of the objects detected in the NIR, 
since it is possible that only  part 
of the energy emitted by the 
stars is reprocessed by the dusty envelope. In that sense, they can be taken 
as lower limits to the bolometric luminosity of the embeded stars.

Assuming that the radio emission measured by Caswell \& Haynes (1987) originates in this region, at a 
distance of 3.7 kpc,
we estimated the number of ionizing Lyman continuum photons  as $N_{Ly} \sim 6 \times 10^{48}$ 
photons s$^{-1}$. On the other hand, the number of
Lyman continuum photons expected from the stars that lie along the reddening line 
for early-type stars
is about 1.7 $ \times$ 
10$^{49}$ s$^{-1}$, enough to produce the detected flux densities at 5 GHz.
The IRS1 source is enough to account for more than 90\% of the total 
number of Lyman continuum
photons necessary to ionize the gas.

Analysis of the integrated  flux densities of the NIR  nebula at the $J$, $H$ and nb$K$ bands
revealed that they increase with frequency, implying that free-free
emission cannot be the main source of the extended luminosity, unless we assume only four magnitudes
of visual extinction. Since this value is incompatible with the extinction  derived
from the stars that do not shown excess of emission at 2.2 $\mu m$, we conclude that the observed
emission must be mainly dust scattered light.

A comparison of the cluster described in this paper with the young stellar cluster associated with the IRAS source
16177-5018 (Roman-Lopes et al. 2003), which is located at the same distance and 
direction, shows that the the former contains  less massive stars. Since its mean visual absorption 
 is also about 10 magnitudes smaller than that of IRAS 16177-5018, 
it is possible that it was formed from a less massive
molecular cloud.

\acknowledgments

This work was partially supported by the Brazilian agencies FAPESP and CNPq.
We acknowledge the staff of Laborat\'orio Nacional de 
Astrof\' \i sica for their efficient support and to Anderson Caproni by
help during the observations.
This publication makes use of data products from the Two Micron All 
Sky Survey, which is a joint project of the University of 
Massachusets and the Infrared Processing and Analysis Center/California 
Institute of Technology, funded by the
National Aeronautics and Space Administration and the National Science Foundation. 
This research made use of data products from the Midcourse Space 
Experiment.

\begin{deluxetable}{crrrrr}
\tabletypesize{\scriptsize}
\tablecaption{Integrated fluxes from MSX $A, C, D$ and $E$ images for
the two mid-infrared sources (see text). Also are shown the data from MSX and IRAS 
point source catalogue ($psc$).  \label{tbl-1}}
\tablewidth{0pt}
\tablehead{
\colhead{$\lambda(\mu$m)}&\colhead{Integration "A"}&\colhead{Integration "B"}&
\colhead{MSX $psc$ "A"}&\colhead{MSX $psc$ "B"}&\colhead{IRAS $psc$}
}
\startdata 

8.28  &15.8   &4.3   &11.3   &4.1   &\nodata\\
12  &\nodata   &\nodata   &\nodata   &\nodata   &44.8\\
12.13  &32.8   &3.6   &23.0   &\nodata   &\nodata\\
14.65  &23.9   &2.0   &21.3   &\nodata   &\nodata\\
21.33  &117   &10.8   &123.8   &\nodata   &\nodata\\
25 &\nodata   &\nodata   &\nodata   &\nodata   &266\\
60  &\nodata   &\nodata   &\nodata   &\nodata   &2311\\
100  &\nodata   &\nodata   &\nodata   &\nodata   &4618\\

\enddata

\end{deluxetable}

\clearpage

\clearpage

\begin{deluxetable}{crrrrrrrrr}
\tabletypesize{\scriptsize}
\tablecaption{List of the selected near-infrared sources \label{tbl-2}}
\tablewidth{0pt}
\tablehead{
\colhead{IRS}&\colhead{$\alpha$(J2000)}&\colhead{$\delta$(J2000)}&
\colhead{$J_{CamIV}$}&\colhead{$J_{2mass}$}&\colhead{$H_{CamIV}$}&
\colhead{$H_{2mass}$}&\colhead{$K_{CamIV}$}&\colhead{$K_{2mass}$}&
\colhead{$Spec$ $Type$}}
\startdata
1 &16:17:02.20 &$-$50:47:03.1 &12.26(4) &11.97(4) &10.88(4) &10.65(4) &10.07(4) & 9.73(4) & O5\\
2 &16:17:09.23 &$-$50:47:14.7 &13.36(3) &13.26(4) &11.66(2) &11.47(3) &10.13(3) &10.15(3) & mid-O\\
3 &16:16:55.8  &$-$50:47:23   &11.77(2) &\nodata      &11.01(2) &\nodata      &10.30(4) &\nodata  & early-B \\
4 &16:17:00.62 &$-$50:47:49.6 &12.47(1)      &\nodata      &12.02(2) &\nodata     &11.17(4) &10.92(2) & early-B\\
5 &16:17:02.63 &$-$50:46:56.6 &14.43(3) &14.26(8) &12.78(3) &12.51(7) &11.64(4) &11.17(7) & early-B\\
6 &16:17:03.7  &$-$50:47:05   &14.61(3) &\nodata      &13.04(2) &\nodata      &12.03(5) &\nodata   & B0  \\
7 &16:17:02.88 &$-$50:47:05.1 &17.08(9) &\nodata      &14.84(6) &\nodata      &11.55(5) &11.62(7) & early-B\\
8 &16:17:06.30 &$-$50:47:09.3 &13.23(2) &13.31(3) &12.28(3) &12.34(3) &11.42(3) &11.70(3) & B0\\
9 &16:17:04.37 &$-$50:47:16.4 &13.28(2) &13.26(2)  &12.44(2) &12.45(3) &11.64(3) &11.78(5) & B0.5\\
10 &16:17:03.72 &$-$50:47:15.7 &13.86(3) &13.92(4) &12.57(3) &12.69(4) &11.80(4) &11.92(5) & B0\\
11 &16:16:56.0 &$-$50:47:22 &13.12(3) &\nodata    &12.62(4) &\nodata    &12.06 &\nodata   & early-B\\
12 &16:16:59.72 &$-$50:47:47.0 &17.32(10) &\nodata    &14.76(6) &\nodata    &12.25(5) &12.31(6) & early-B\\
13 &16:17:02.06 &$-$50:47:26.6 &16.27(6) &\nodata    &13.95(6) &\nodata    &12.54(5) &12.60(9) & O9.5\\
14 &16:17:08.61 &$-$50:47:11.5 &15.39(5) &\nodata    &13.76(3) &13.65(6) &12.64(9) &12.65(2) & B0.5\\
15 &16:17:02.0 &$-$50:47:01 &16.23(6) &\nodata    &14.35(4) &\nodata    &12.51(6) &\nodata  & early-B \\
16 &16:17:04.16 &$-$50:47:49.8 &13.89(3) &\nodata    &13.27(3) &\nodata    &12.87(6) &12.72(7) & B2\\
17 &16:17:03.13 &$-$50:47:15.4 &14.74(3) &14.61(6) &13.65(3) &13.57(5) &12.71(6) &12.73(7) & early-B\\
18 &16:16:55.6 &$-$50:47:28 &15.74(5) &\nodata    &14.24(4) &\nodata    &12.61(6) &\nodata   & early-B\\
19 &16:17:03.7 &$-$50:47:08 &15.53(5) &\nodata    &13.91(4) &\nodata    &12.69(7) &\nodata   & early-B\\
20 &16:17:02.02 &$-$50:48:12.1 &14.94(3) &14.90(5) &13.89(4) &13.71(5) &12.55(7) &12.88(5) & B2\\
21 &16:16:55.3 &$-$50:47:25 &16.22(8) &\nodata    &14.56(5) &\nodata    &12.95(6) &\nodata   & early-B\\
22 &16:17:02.5 &$-$50:46:54 &16.35(8) &\nodata    &14.47(5) &\nodata    &12.98(6) &\nodata   & early-B\\
23 &16:17:03.93 &$-$50:47:34.0 &16.50(7) &\nodata    &14.42(4) &14.42(8) &13.31(7) &13.20(5) & B0.5\\
24 &16:17:03.9 &$-$50:47:13 &18.17(18) &\nodata    &15.16(7) &\nodata    &13.12(6) &\nodata   & early-B\\
25 &16:16:59.41 &$-$50:47:52.9 &17.16(9) &\nodata    &15.04(5) &\nodata    &13.11(6) &13.31(7) & early-B\\
26 &16:17:08.12 &$-$50:47:37.4 &16.21(6) &\nodata    &14.47(4) &14.51(7) &13.29(8) &13.38(6) & B2\\
27 &16:17:02.12 &$-$50:47:37.2 &16.54(8) &\nodata    &14.50(4) &\nodata    &13.42(7) &13.40(11) & B1\\
28 &16:17:03.8 &$-$50:47:11 &17.40(11) &\nodata    &15.36(6) &\nodata    &13.32(7) &\nodata   & early-B\\
29 &16:17:02.2 &$-$50:47:35 &16.63(7) &\nodata    &14.86(7) &\nodata    &13.61(7) &\nodata   & early-B\\
30 &16:17:03.88 &$-$50:48:16.0 &16.08(7) &15.59(9) &15.54(7) &14.69(9) &\nodata    &13.74(8) & mid-B\\
31 &16:17:05.23 &$-$50:47:31.7 &15.28(4) &15.20(9) &14.63(5) &\nodata    &13.53(5) &13.87(8) &\nodata\\
32 &16:17:06.74 &$-$50:48:05.2 &16.16(7) &15.85(8) &15.37(6) &\nodata    &\nodata    &14.18(10) &\nodata\\
33 &16:16:55.9 &$-$50:47:20 &13.62(5) &\nodata    &12.93(4) &\nodata   &11.90(4) &\nodata   & early-B\\
34 &16:17:03.8 &$-$50:47:47 &16.42(6) &\nodata   &14.25(5) &\nodata    &12.85(6) &\nodata   & B0\\
35 &16:17:03.5 &$-$50:47:03 &15.67(5) &\nodata   &13.79(4) &\nodata    &12.56(6) &\nodata   & B0.5\\
\enddata


\tablenotetext{* }{Stars that show "excess" of emission at 2.2 $\mu$m }

\end{deluxetable}

\clearpage

\begin{figure*}
\epsscale{2.2}\plotone{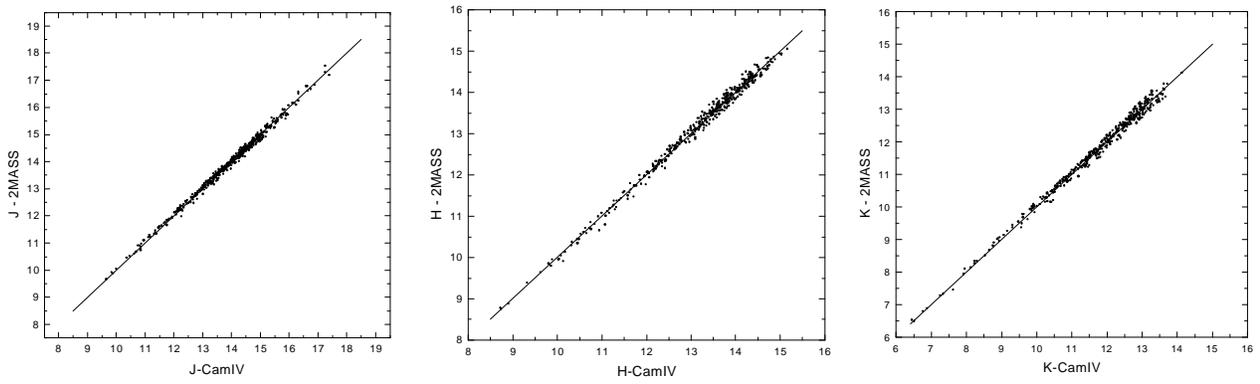}
\caption{Magnitudes comparative diagram M$_{JHK}$(2MASS) $\times$ M$_{JHK}(CamIV)$. The continuous line shows the relation expected
if the two photometric systems were equal.  \label{fig1}}
\end{figure*}

\clearpage

\begin{figure*}
{\plotone{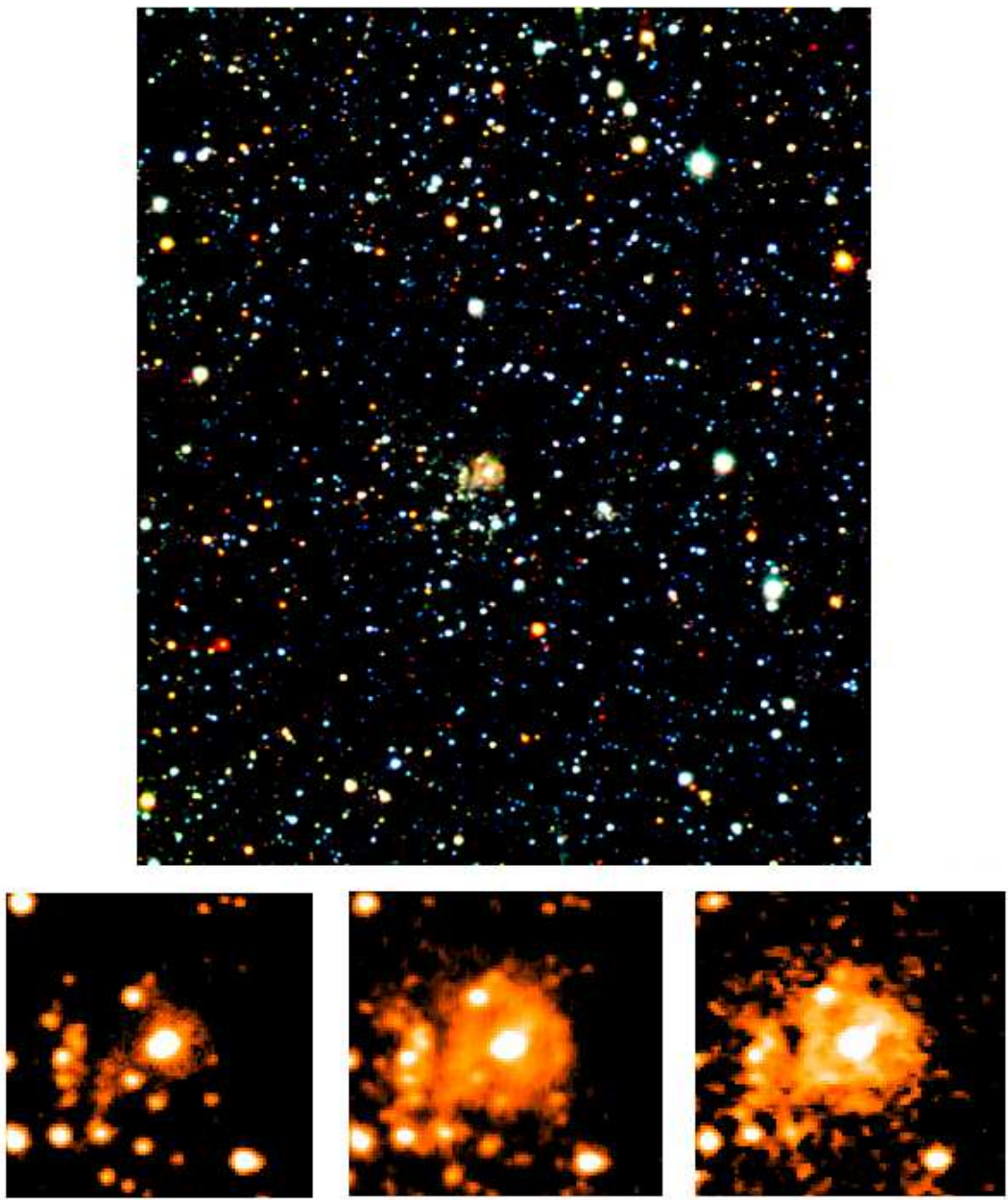}}
\caption{Combined false-image made from the $J, H$ and nb$K$ LNA images. North is to the top east to 
the right. At the botton are detailed views of the infrared nebulae 
$J (left), H (center)$ and nb$K (right)$ images.\label{fig2}}
\end{figure*}

\begin{figure*}
{\epsscale{2}\plotone{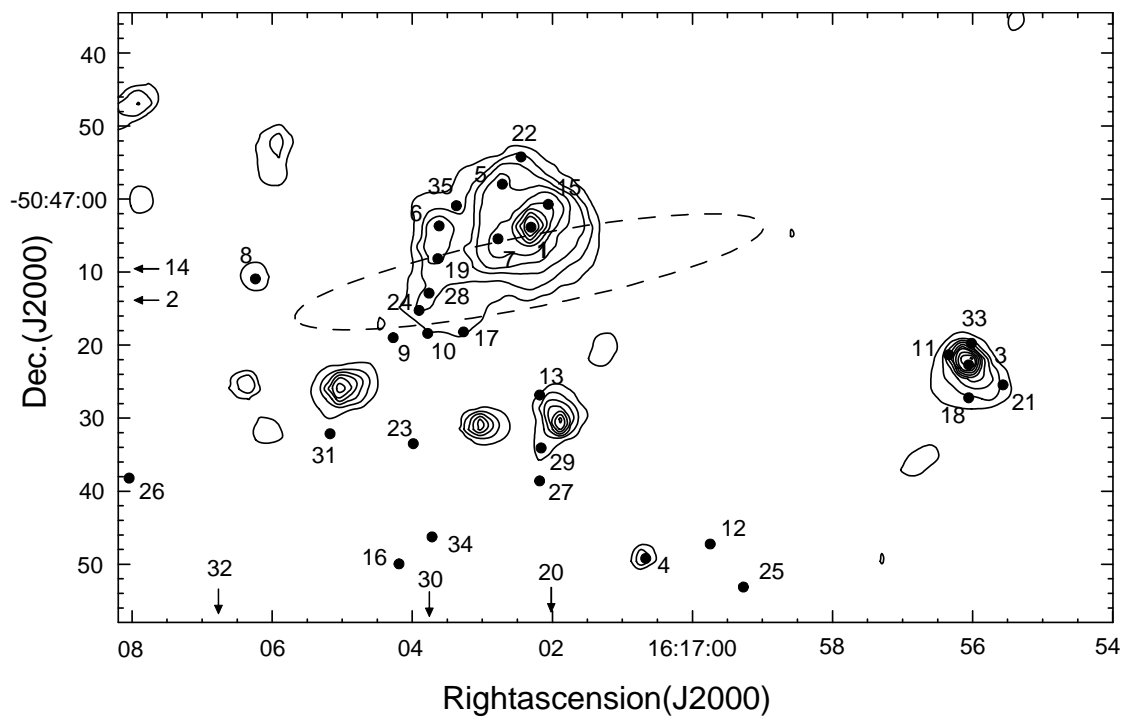}}
\caption{nb$K$ band contour map of the infrared nebulae associated with IRAS 16132-5039.
The contours start at 2.2$\times 10^{-4}$ Jy/beam, with the same intervals (the beam size is $2 \times 2$ \ pixels).
The positions of selected infrared sources refered in Table 2 and the IRAS coordinate elipse error 
(dotted line) also are indicated.\label{fig3}}
\end{figure*}

\clearpage

\begin{figure*}
{\plotone{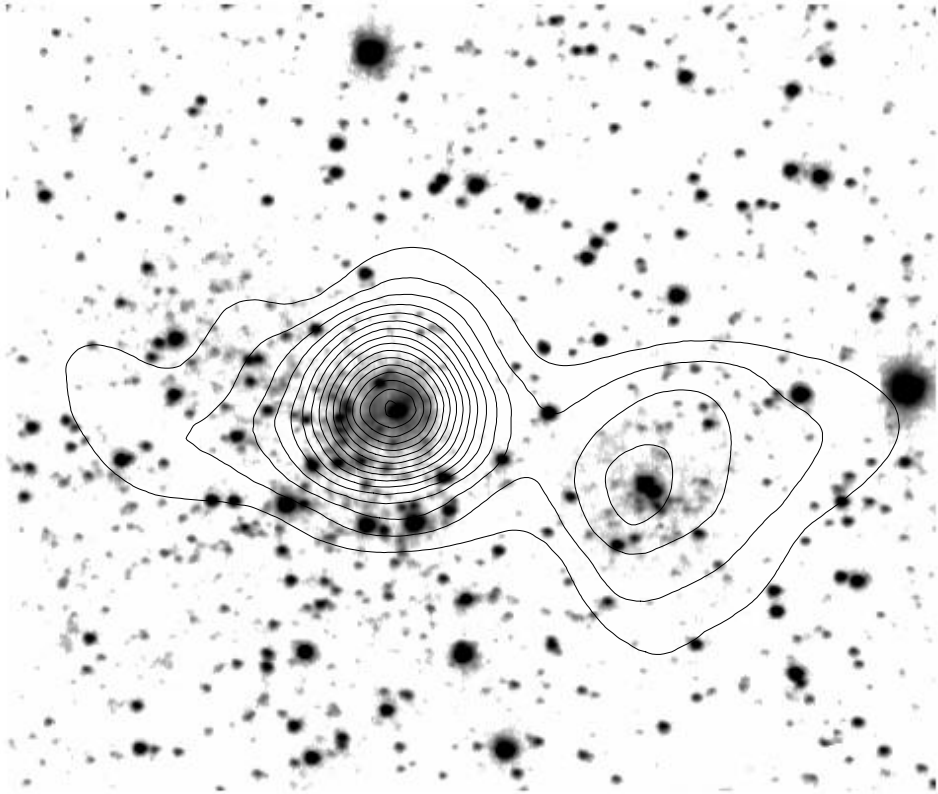}}
\caption{Contour diagram from the $A$ MSX band image (8.28$\mu$m), overlaying a LNA's $H$ band image.
The contours start at 2.8$\times 10^{-5}$ W/$m^{2}$sr and are spaced by 8$\times 10^{-6}$ W/$m^{2}$sr.\label{fig4}}
\end{figure*}

\begin{figure*}
{\plotone{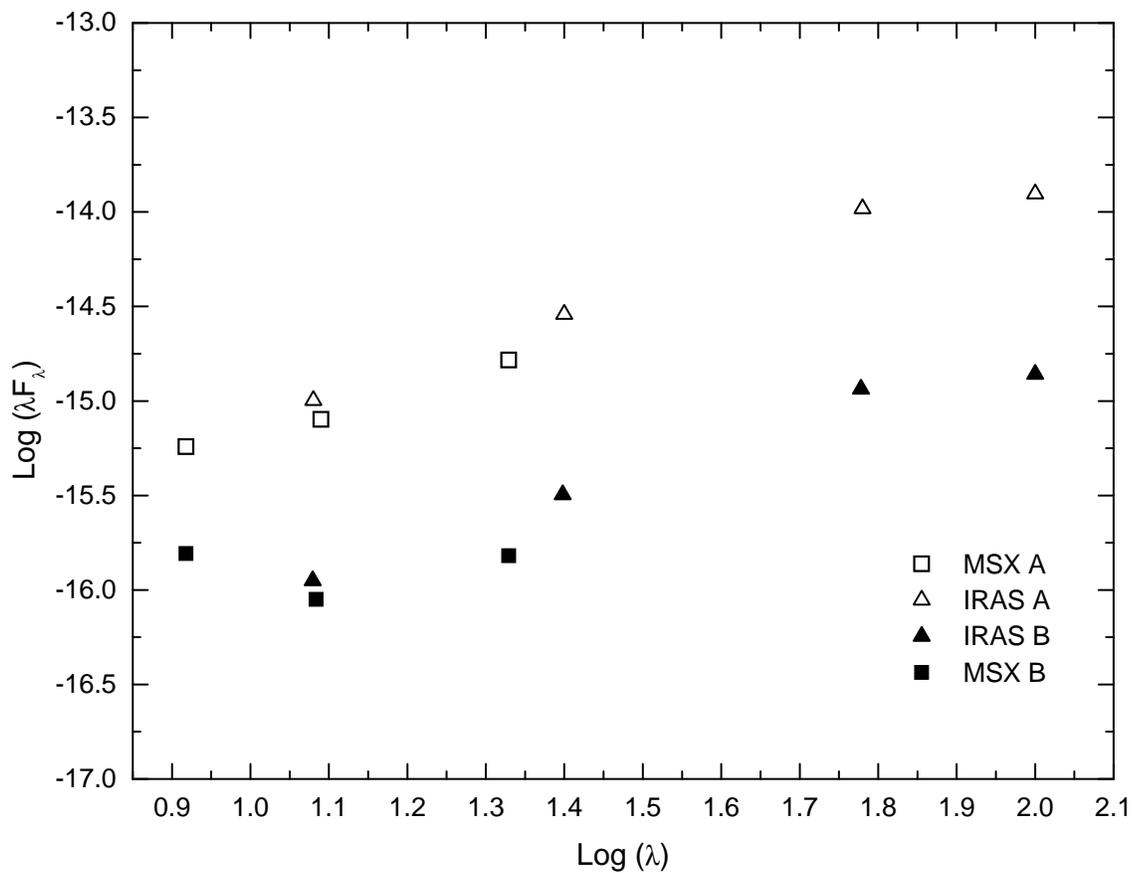}}
\caption{Spectral energy distribution of the "A" and "B" MSX sources. The mid infrared data (squares) were 
obtained from the integrated flux of the individual sources in the MSX images
(bands A=8.28$\mu$m, C=12.13$\mu$m, D=14.65$\mu$m and E=21.34$\mu$m) while the far infrared data 
(triangles) were taken from IRAS, and are listed in Table 1.\label{fig5}}
\end{figure*}

\begin{figure*}
{\plotone{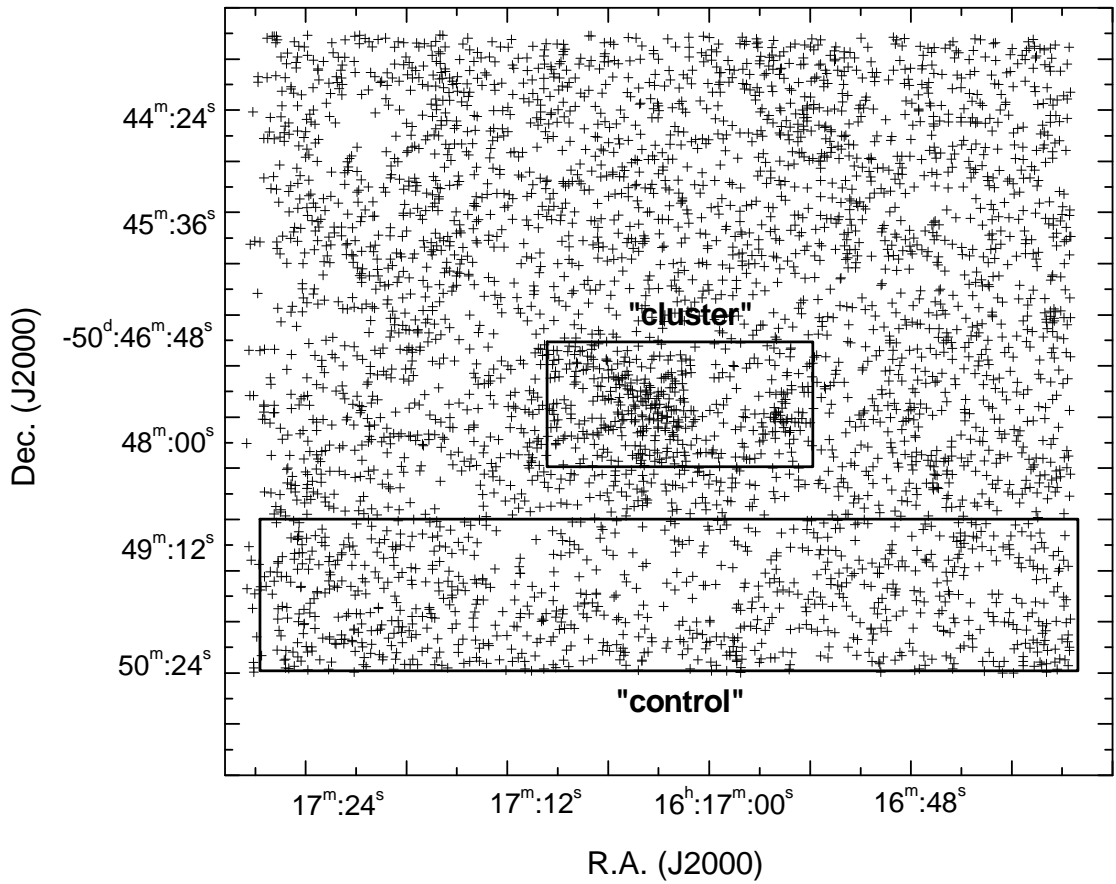}}
\caption{Diagram of the spacial distribution of sources detected at LNA's $\it{H}$ band image. 
The two regions
({\bf a}) (cluster) and ({\bf b}) (control) are delimitated by boxes .\label{fig6}}
\end{figure*}

\clearpage

\begin{figure*}
{\includegraphics{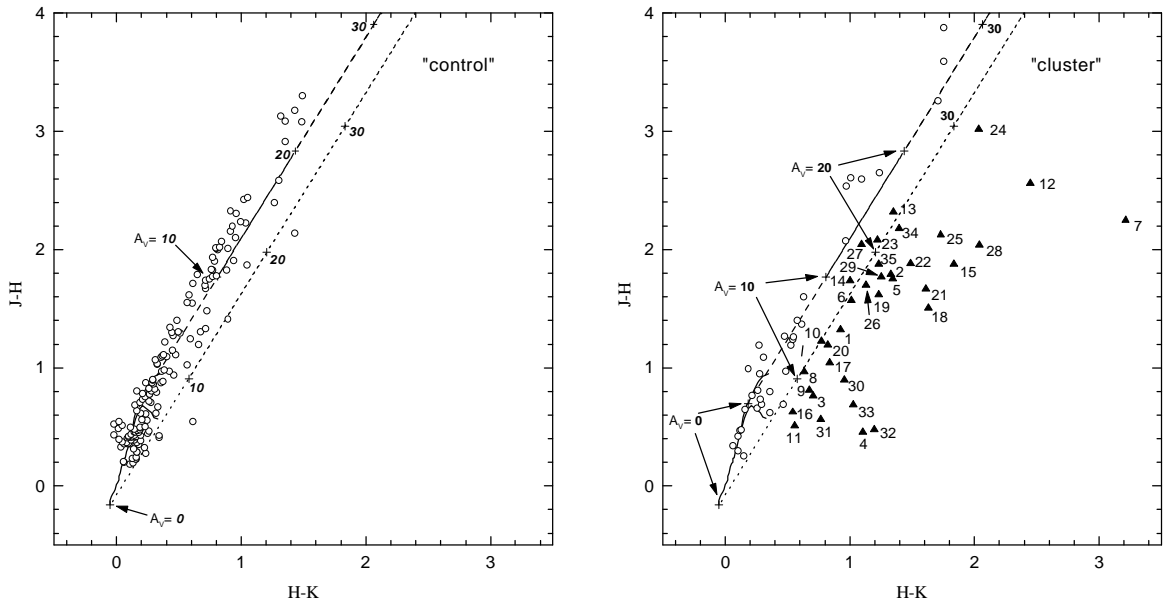}}
\caption{Color-color diagrams for two regions in our survey. The region labeled "control" (left panel) contains
only foreground objects; the other labeled "cluster" (right panel) has also a foreground population (open circles)
but presents 
objects with infrared "excess" (filled triangles). All  objects with "excess"
are located in the "cluster area". The locus of the main sequence
and giants branch are shown by the continuous lines taken from Koornneef (1983),
while the dashed and dotted lines follow the reddening vectors taken from Rieke $\&$ Lebofsky (1985).
The location (plus signs) of $A_V$ = 0, 10, 20 magnitudes of visual extinction also are indicated.
The cluster members candidates are labeled by numbers.\label{fig7}}
\end{figure*}

\begin{figure*}
{\epsscale{1.5}\plotone{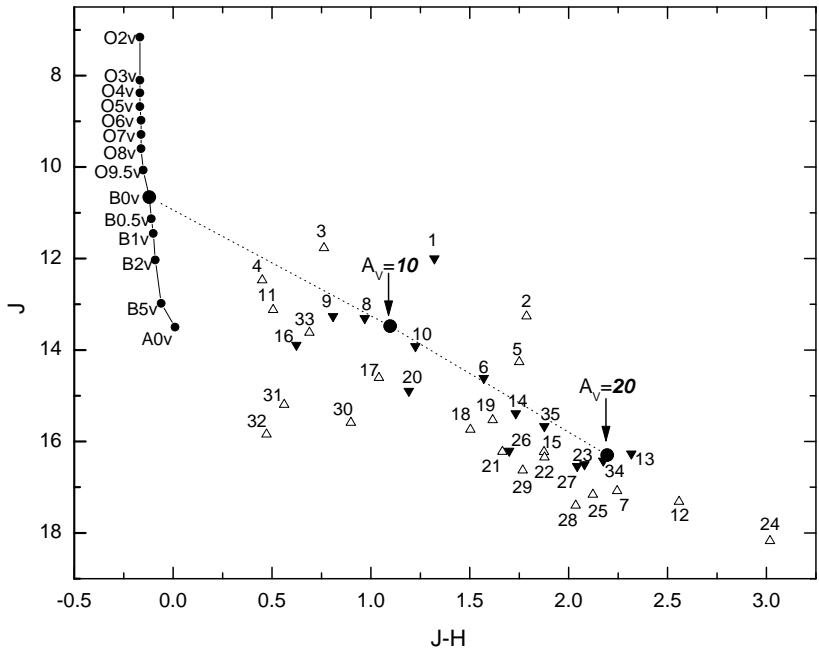}}
\caption{The $J$ versus $(J-H)$ color-magnitude diagram of the sources in table 2.
The locus of the main sequence at 3.7 kpc is shown by the continuous line. The intrinsic colors were taken
from Koornneef (1983) while the absolute $\it{J}$ magnitudes were calculated from the absolute visual luminosity for ZAMS taken from
Hanson et al. (1997). The reddening vector for a B0 ZAMS star (dotted line) was taken
from Rieke $\&$ Lebofsky (1985). We also indicated  the location (bold numbers) of $A_V$ = 10 and 20 
magnitudes of visual extinction as well as the sources that show "excess" (open up triangles) 
and do not (filled down triangles) in the color-color diagram.\label{fig8}}
\end{figure*}

\clearpage

\begin{figure*}
{\epsscale{1}\plotone{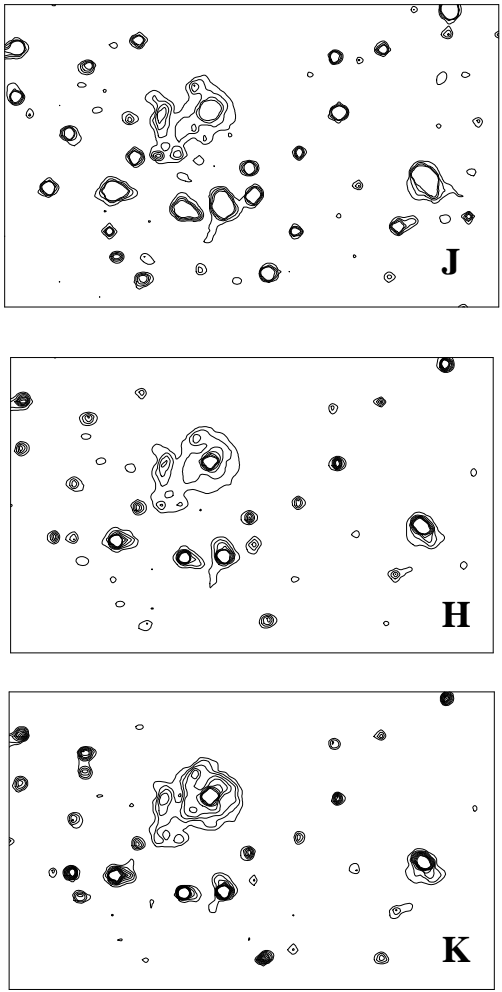}}
\caption{ The $J, H$ and nb$K$ contour maps of the infrare nebula region. The contours start at 
0.44 ($J$), 1.6 ($H$) and 
2.2$\times 10^{-4}$ Jy/beam ($K$), and are spaced by 0.37, 1.8 and 1.1$\times
10^{-4}$ Jy/beam respectively.\label{fig9}}
\end{figure*}

\end{document}